\documentclass[conference]{IEEEtran}
\IEEEoverridecommandlockouts

\usepackage{cite}
\usepackage{amsmath,amssymb,amsfonts}
\usepackage{algorithmic}
\usepackage{graphicx}
\usepackage{textcomp}
\usepackage{tcolorbox}
\usepackage{xcolor}
\usepackage{url}
\usepackage{multirow}
\def\BibTeX{{\rm B\kern-.05em{\sc i\kern-.025em b}\kern-.08em
    T\kern-.1667em\lower.7ex\hbox{E}\kern-.125emX}}

\begin{document}

\title{Agentic Pipelines in Embedded Software Engineering: Emerging Practices and Challenges
\thanks{This paper has been partially financed by Software Center,
www.software-center.se, a collaboration between Chalmers, the University of Gothenburg, and 17 companies.}
}

\author{\IEEEauthorblockN{Simin Sun}
\IEEEauthorblockA{\textit{Chalmers University of Technology} \\
\textit{and University of Gothenburg}\\
Gothenburg, Sweden \\
simin.sun@gu.se}
\and
\IEEEauthorblockN{Miroslaw Staron}
\IEEEauthorblockA{\textit{Chalmers University of Technology} \\
\textit{and University of Gothenburg}\\
Gothenburg, Sweden \\
miroslaw.staron@gu.se}
}

\maketitle

\begin{abstract}
A new transformation is underway in software engineering, driven by the rapid adoption of generative AI in development workflows. Similar to how version control systems once automated manual coordination, AI tools are now beginning to automate many aspects of programming. For embedded software engineering organizations, however, this marks their first experience integrating AI into safety-critical and resource-constrained environments. The strict demands for determinism, reliability, and traceability pose unique challenges for adopting generative technologies.

In this paper, we present findings from a qualitative study with ten senior experts from four companies who are evaluating generative AI-augmented development for embedded software. Through semi-structured focus group interviews and structured brainstorming sessions, we identified eleven emerging practices and fourteen challenges related to the orchestration, responsible governance, and sustainable adoption of generative AI tools. Our results show how embedded software engineering teams are rethinking workflows, roles, and toolchains to enable a sustainable transition toward agentic pipelines and generative AI-augmented development.

\end{abstract}


\section{Introduction}

During the early days of computing, software development was labor-intensive and required large teams to write, test, and debug code manually. As tools, frameworks, and automation evolved, especially with the rise of Integrated Development Environments (IDEs), Continuous Integration/Continuous Deployment/Delivery (CI/CD), and generative AI-augmented coding assistants, the number of developers needed for many tasks has decreased. Today, a single developer can leverage advanced tools to build, test, and deploy complex applications individually or in small teams, accomplishing what used to demand entire programming departments.

Much like version control and automated code management, generative AI is beginning to automate many programming processes~\cite{terragni2025future}. This raises questions for software engineers about our profession: \emph{How will traditional roles, including developers, testers, and project managers, evolve as generative AI becomes an active participant in the development process?} And \emph{how can organizations adapt their processes and governance to ensure safety, quality, and accountability in this transition?} These questions are crucial for organizations involved in embedded software engineering because deterministic behavior is a fundamental requirement~\cite{lee2025timing}, and LLMs cannot replace programmers but must complement them. As embedded systems grow in size and complexity, reliability~\cite{narayanan2006reliability} and traceability concerns~\cite{steghofer2019software} have also increased, and these concerns have become essential in safety-critical domains such as the automotive industry.

\begin{figure}[!htbp]
    \centering
    \includegraphics[width=0.95\linewidth]{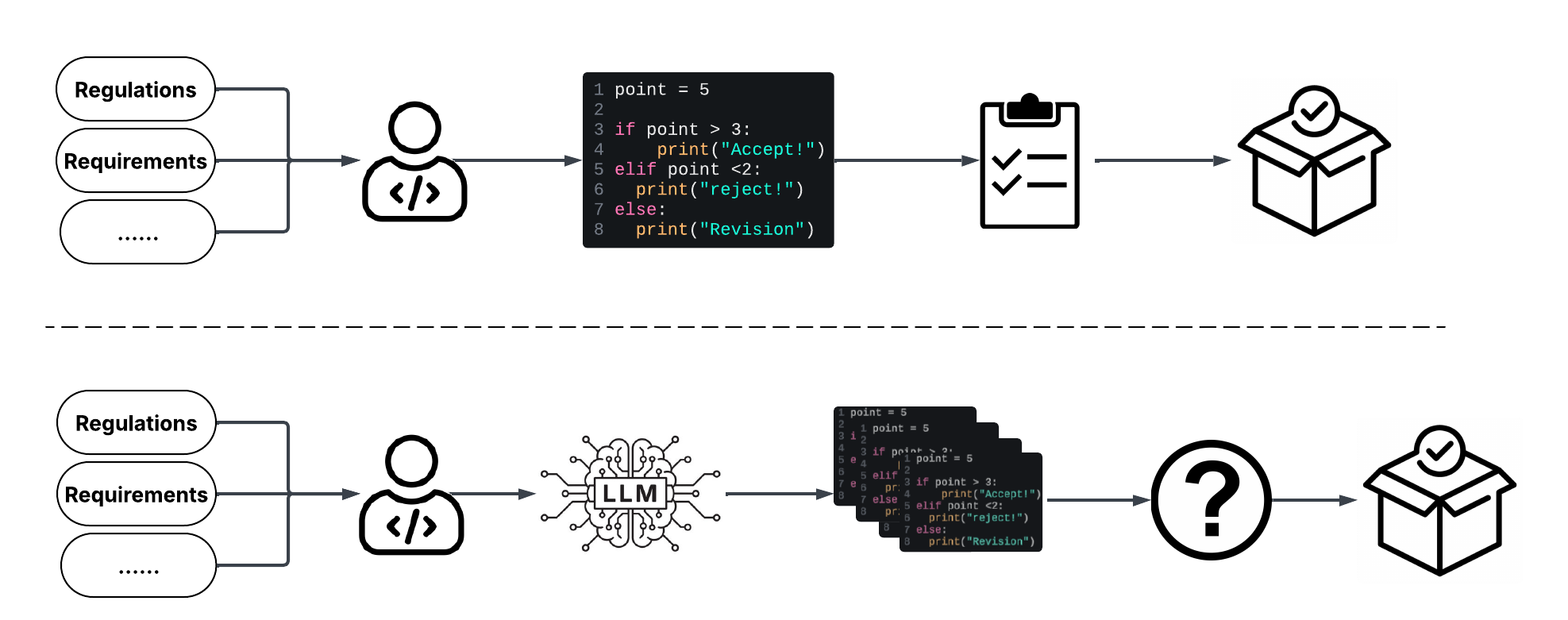}
    \caption{Deterministic (top) vs. probabilistic (bottom) development pipelines.}
    \label{fig:pipeline}
\end{figure}

Given these characteristics, the integration of generative AI into embedded system development presents distinctive challenges. Embedded products must maintain strict determinism, reliability, and traceability, but the development process becomes less predictable when code, tests, and documentation are produced with the help of LLMs, whose outputs can vary across runs or model versions. As illustrated in Fig.~\ref{fig:pipeline}, the traditional pipeline (top) assumes that a fixed set of requirements, regulations, and tools yields a stable, auditable path from specification to implementation, whereas the AI-augmented pipeline (bottom) introduces a probabilistic generative step between intent and artifact. Introducing generative AI–augmented workflows therefore requires not only technical innovation but also new mechanisms for accountability, validation, and certification, so that the resulting artifacts remain auditable and trustworthy even if the path by which they are produced is probabilistic.

While previous studies have described how industrial organizations organize their “AI journeys”~\cite{bosch2025towards, parnin2025building, nahar2025beyond}, much less is known about how embedded development practices evolve as generative AI tools become part of their workflows where requirements on determinism, reliability, and traceability are non-negotiable. In particular, we lack empirical accounts of how practitioners in these settings interpret, adapt to, and govern the introduction of generative AI in day-to-day engineering work.

This study aims to address this gap by studying the following research question:
\textbf{How are embedded software engineering organizations adopting and adapting to generative AI while maintaining high assurance, and what implications does this have for engineering practices, and governance structures?}

To answer this question, we conducted a qualitative study with ten senior engineers and technical experts from four companies actively exploring the feasibility of generative AI-augmented development in embedded systems. Through focus group interviews and structured brainstorming sessions, we captured practitioners’ firsthand perspectives on the opportunities and challenges of integrating generative AI. By analyzing their experiences, this study contributes empirical insights into how embedded software engineering teams navigate determinism, reliability, and traceability in the presence of probabilistic AI systems. The findings not only highlight emerging practices, such as human-in-the-loop supervision, generative AI-friendly artifact design, and compiler-in-the-loop feedback, but also illustrate how organizations are preparing for a sustainable, auditable transition toward generative AI-augmented engineering.

The rest of the paper is structured as follows:
Section~\ref{sec:background} presents background on embedded systems and generative AI tools. Section~\ref{sec:methodology} describes our research design. Section~\ref{sec:findings} presents the key findings, and Section~\ref{sec:discussion} discusses their implications for future software engineering practices. Finally, Section~\ref{sec:conclusion} concludes the paper.

\section{Background}
\label{sec:background}
\subsection{Embedded Software Engineering}
Embedded systems are specialized computing units that integrate software and hardware to perform dedicated functions under strict resource, timing, and safety constraints~\cite{wolf2012computers}. They are foundational to critical domains such as automotive, health care, and aerospace, where software failures can lead to safety hazards or financial loss~\cite{solouki2024dependability}. Typical embedded architectures require real-time execution, deterministic behavior, and tight coupling with hardware interfaces~\cite{olsson2020going}. Development processes are therefore guided by rigorous standards such as IEC-61508~\cite{IEC61508_2010_CMV} for general machinery, ISO-26262~\cite{ISO26262_2018} for automotive, and DO-178C~\cite{RTCA_DO178C_2011} for aerospace.

These properties make embedded systems engineering inherently disciplined, prioritizing predictability, reproducibility, and traceability throughout the development lifecycle. At the same time, growing software complexity, coupled with shorter release cycles, is pushing the field toward new automation techniques. Recent studies have highlighted early explorations of generative AI in embedded contexts, such as real-time scheduling for generative AI workloads~\cite{lee2025timing}, and fault prediction in safety-critical systems~\cite{solouki2024dependability}. However, integrating generative AI into embedded workflows raises new challenges related to certification, version control, and determinism~\cite{kellogg2020continuous}. These characteristics make embedded systems a compelling testbed for studying how generative AI can coexist with regulated, high-assurance development environments.

\subsection{Generative AI in SE}
Recent advances in LLMs, including GPT-5, Sonnet 4.5, Claude 3, and Gemini 2, have demonstrated broad applicability to software engineering tasks such as automated code completion, documentation generation, bug fixing, and test synthesis~\cite{nahar2023meta, parnin2025building}. Tools like GitHub Copilot, Amazon CodeWhisperer, and Meta’s CodeCompose are now routinely used in industrial software development~\cite{murali2023codecompose, vaithilingam2022expectation}.  

In research, LLMs are increasingly studied as components of \textit{agentic software engineering} frameworks, where multiple generative AI agents collaborate across tasks such as requirements refinement, software development and testing, and maintenance~\cite{terragni2025future}. However, despite rapid progress, scholars identify persistent limitations, including non-determinism, inconsistent reasoning, and explainability deficits~\cite{hassan2024rethinking}. These issues challenge core principles of embedded development: verification, traceability, and reproducibility. Addressing them requires both technical solutions and new organizational practices for responsible adoption~\cite{staron2025exploring}.

Integrating generative AI into embedded systems engineering introduces both opportunities and risks. On the one hand, generative AI can automate code generation, documentation, and testing for peripheral drivers or configuration files, significantly accelerating prototyping and maintenance. On the other hand, the non-deterministic behavior of LLMs~\cite{ouyang2025empirical}, the lack of transparent provenance, and the evolution of model versions raise concerns about long-term support and the validity of certification~\cite{kellogg2020continuous}.

\section{Methodology} 
~\label{sec:methodology}
To explore the challenges and opportunities of integrating generative AI into software engineering practices, we conducted a mixed-methods study involving semi-structured focus group interviews and structured brainstorming sessions. The interviews provided in-depth insights into practitioners’ daily challenges and strategies, while the brainstorming sessions enabled participants to articulate future directions and shared concerns collaboratively. Combining both methods, the study aimed to understand the current state of generative AI adoption in software development as well as participants’ visions for its evolution over the next five years. 

We first conducted semi-structured focus group interviews with ten software engineers who are actively exploring the feasibility and impact of generative AI technologies in their organizations. Subsequently, we facilitated Mentimeter-based brainstorming sessions with the same participants to discuss anticipated developments over the next five years. These sessions included discussions focused on how generative AI could transform software engineering paradigms at multiple levels: individual roles, tools, team dynamics, organizational strategies, and investment priorities.

\subsection{Participants}
Participants were recruited through targeted invitations to senior professionals engaged in generative AI initiatives within their organizations. We used purposeful sampling to ensure that all ten participants from four companies had relevant roles, experience, and firsthand involvement in generative AI adoption.
The participating companies varied in size and industry domain, as shown in Table~\ref{tab:demographics}. For classification, we defined company scales as follows: Small (\textless 1,000 employees), Medium (1,000–10,000 employees), and Large (\textgreater 10,000 employees).

\begin{table}[htbp]
\caption{Demographic information about participating companies}
\centering
\begin{tabular}{lll} \hline
Company & Domain & Scale \\ \hline
C1      & Communications       &  Medium     \\ 
C2      & Electronics          &  Large      \\ 
C3      & Energy Control       &  Small       \\
C4      & Machinery  &  Large       \\ \hline 
\end{tabular}
\label{tab:demographics}
\end{table}

\subsection{Procedure}
\subsubsection{Focus-group Interview} 
The focus group interviews followed a semi-structured format designed to balance open discussion with a consistent inquiry framework across sessions. Each session began with participant introductions, followed by brief presentations from each company outlining their organizational background and current generative AI initiatives. 

Participants were then guided through a series of prompting questions covering three key themes:
\begin{enumerate}
    \item the current state of generative AI use within their organization;
    \item their individual and team experiences with adopting and applying generative AI tools in daily development practices;
    \item their concerns, challenges, and expectations regarding the future integration of generative AI in software engineering.
\end{enumerate}
To facilitate discussion, participants were encouraged to share specific examples, reflect on lessons learned, and identify both opportunities and risks observed in practice.

\subsubsection{Mentimeter-based brainstorming} Following the interviews, participants engaged in a structured brainstorming session facilitated through Mentimeter. This interactive tool supported real-time idea collection, clustering, and voting, helping the group articulate collective insights about the expected evolution of generative AI in software engineering practices, tools, and roles over the next five years. We included the questions and their types in Table~\ref{tab:menti}.

\begin{table}[htbp]
\caption{List of Mentimeter questions}
\centering
\scriptsize
\label{tab:menti}
\begin{tabular}{p{7cm} p{1cm}} \hline
\textbf{Question} & \textbf{Type} \\ \hline

Which technology will have the most impact:\newline
- Agentic AI (SE 4.0)\newline
- Programming assistants (like CoPilot)\newline
- Prompt engineering (Software 3.0)\newline
- CI/CD assistants\newline
- Intelligent test assistants\newline
- Code review assistants
& 100 Points \\ \hline

Which roles will need to adapt the most (re-skilling):\newline
- Programmers\newline
- Architects and designers\newline
- Testers\newline
- UX designers\newline
- Requirement engineers\newline
- Project managers\newline
- Release engineers
& 100 Points \\ \hline

How important is it to build competence in-house for:\newline
- Integrating AI into products\newline
- Prompt engineering\newline
- Developing agents\newline
- Using AI for SE\newline
- AI engineering\newline
- Software-on-demand
& 100 Points \\ \hline

What hinders the adoption of these tools?
& Open End \\ \hline

If you were to invest in improving AI, what would you improve?\newline
- Transparency in generation\newline
- Oracles for validation\newline
- Hallucinations\newline
- Speed\newline
- Data leakage (from prompts/tools)\newline
- Termination criteria\newline
- Time frame for solving tasks
& 100 Points \\ \hline

If you were to invest in your organization, what would it be?\newline
- New competence\newline
- New tools / better tools\newline
- Integration with current tools\newline
- New models / better models\newline
- AI-friendly code\newline
- Refactoring / rewriting legacy software\newline
- Dedicated AI team
& 100 Points \\ \hline

What will happen to your team in the future (3–5 years)?\newline
- Agentic AI\newline
- AI support\newline
- Cross functionality\newline
- Number of people\newline
- Interaction with external stakeholders\newline
- Typical SE roles\newline
- Programming done by non-programmers
& Scales \\ \hline

What will happen to your organization (3–5 years)?\newline
- SE roles\newline
- AI roles\newline
- Sales roles\newline
- Product management\newline
- Administration (HR, managers)\newline
- Domain specialists\newline
- Non-SE roles doing software development\newline
- New / diverse products
& Scales \\ \hline

What will happen to software engineering (3–5 years)?\newline
- Programmers\newline
- Project managers\newline
- Requirements engineers\newline
- UX designers\newline
- Architects / designers\newline
- Testers\newline
- Product managers\newline
- Process owners / managers
& Scales \\ \hline

Where would you like AI/AI agents to have the most impact (3–5 years)?\newline
- Programming\newline
- Compliance\newline
- Maintenance\newline
- Testing and QA\newline
- Design and refactoring\newline
- Code review\newline
- Program repair\newline
- Onboarding
& 100 Points \\ \hline

When can the human-in-the-loop be skipped?
& Open Ended \\ \hline

How to validate and verify the system with the generated code?
& Open Ended \\ \hline

What about long-term support and traceability in 10 years?
& Open Ended \\ \hline

\end{tabular}
\end{table}

Both Focus-group interviews and brainstorming sessions were transcribed and analyzed using thematic analysis conducted by two researchers. This approach enabled systematic identification and interpretation of recurring themes and patterns within the data, capturing participants’ emerging practices and challenges.

\section{Findings}
~\label{sec:findings}

\noindent

\begin{figure}[htbp]
    \centering
    \includegraphics[width=\linewidth]{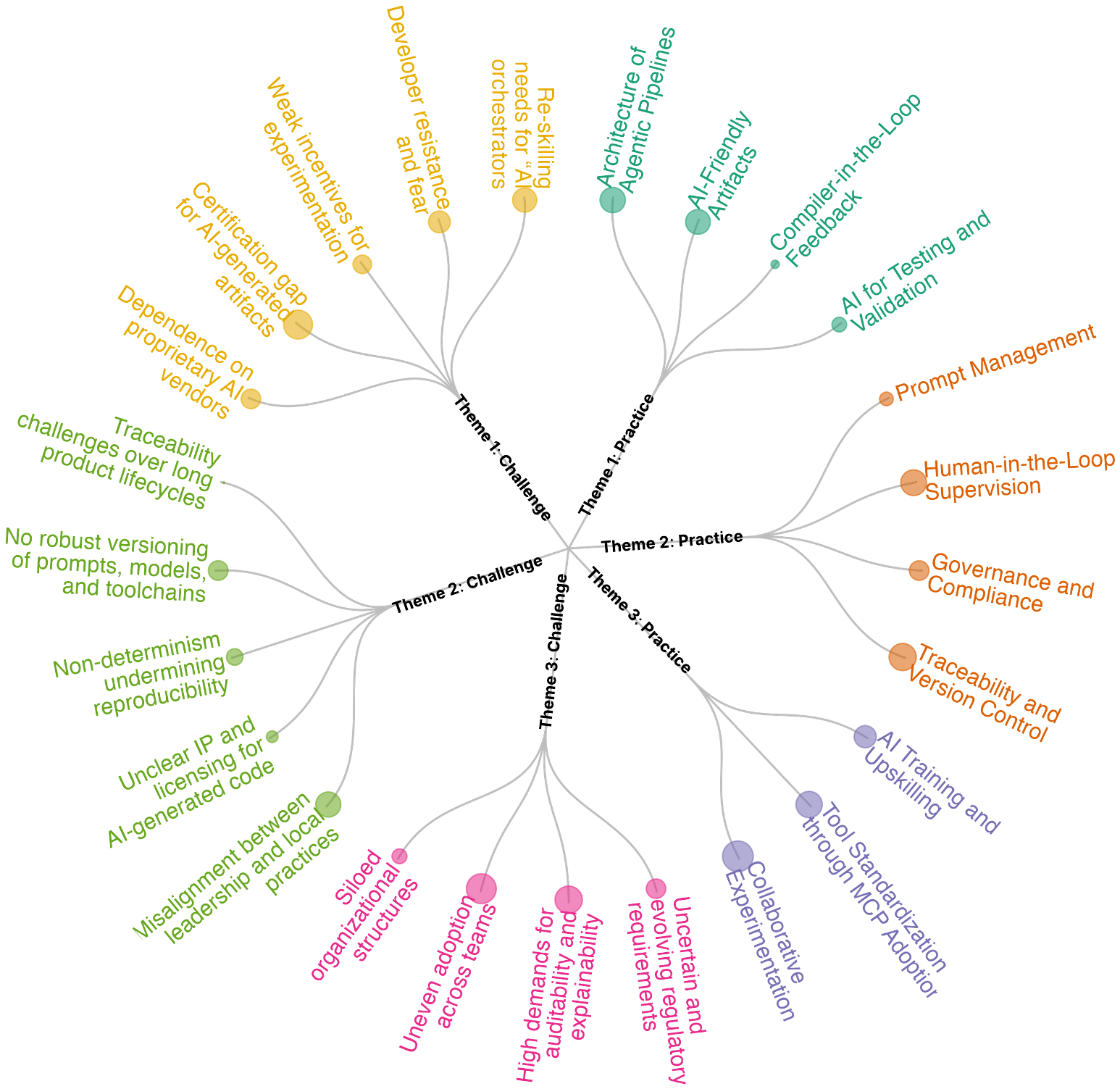}
    \caption{Clusters of emerging practices and challenges into three themes.}
    \label{fig:dendogram}
\end{figure}

Our analysis shows that the participating organizations are both developing new practices and encountering recurring challenges as they adopt generative AI (see Fig.~\ref{fig:dendogram}). Together, these findings cluster into three themes:

\begin{itemize}
\item \textbf{Theme 1:} Orchestrated AI Workflow: the technical integration of generative AI into engineering processes;
\item \textbf{Theme 2:} Responsible AI Governance: mechanisms for oversight, accountability, and traceability;
\item \textbf{Theme 3:} Sustainable AI Adoption: how companies build and maintain the human and infrastructural capacity needed to use AI at scale.
\end{itemize}

\subsection{Emerging practices}

Table~\ref{tab:practices} summarizes detailed practices and their supporting evidence.

\begin{table*}[htbp]
\caption{Emerging Practices in Adopting Agentic Pipelines for Embedded Systems}
\centering
\begin{tabular}{p{1cm} p{2.5cm} p{5cm} p{6cm} p{1cm}}
\hline
\textbf{Theme} & \textbf{Practice} & \textbf{Description} & \textbf{Supporting Quotes} & \textbf{Participant} \\ 
\hline

\multirow{4}{*}{Theme 1} 
& Architecture of Agentic Pipelines &
Implement multi-agentic CI/CD pipelines integrating generative AI into development, testing, and validation. &
C2: “Now moving towards a multi-agentic way of development.” \newline
C2: “Agents should work in pipelines, like the teams today.” & C2 \\ \cline{2-5}

& AI-Friendly Artifacts &
Adopt AI-friendly code and documentation structures (e.g., \textit{agent.md}, instruction files) for better machine readability. &
C3: “We use instruction files as a writing tool, which is a strong addition to what we do.” \newline
C2: “Use md-files for structuring AI-friendly code.” \newline
C4: “AI-friendly code basically means more structured comments and clear logic.” & C2, C3, C4 \\ \cline{2-5}

& Compiler-in-the-Loop Feedback &
Integrate compiler or simulator feedback loops so AI agents can iteratively correct code errors. &
C2: “We observed that if you just give the LLM access to the compiler, it becomes much better.” \newline
C3: “We use Copilot together with Visual Studio Code in agent mode.” & C2, C3 \\ \cline{2-5}

& AI for Testing and Validation &
Use AI to generate and validate unit tests, compliance checks, and automated documentation for embedded code. &
C3: “We use custom agents for unit testing -- it is a strong addition.” \newline
C1: “Generating test cases is also used a lot.” & C1, C3 \\ \hline

\multirow{4}{*}{Theme 2}
& Prompt Management &
Standardize prompts using version-controlled instruction sets or prompt templates to ensure consistency and compliance. &
C3: “Used instructions, shared among all developers, to make sure consistent styles.” \newline
C2: “Structure the system prompt to reduce hallucinations.” & C2, C3 \\ \cline{2-5}

& Human-in-the-Loop Supervision &
Maintain human oversight during AI-augmented development, with engineers validating generated outputs. &
C2: “Human-in-the-loop works best when the engineer keeps control of the method.” \newline 
C4: “Assistant to assist the developer -- cannot just take the code, needs to rewrite it.” & C2, C4 \\ \cline{2-5}

& Governance and Compliance &
Apply governance frameworks for licensing, compliance checks, and Responsible AI auditing before deployment. &
C1: “We need to know why we need to adopt AI.” \newline
C2: “We're pushing all of our tools have to be in our MCP, have an MCP API as well, so these agents don't have any limitations.” \newline
C3: “License considerations -- who owns the copyright of the generated code.” & C1, C2, C3 \\ \cline{2-5}

& Traceability and Version Control &
Implement traceable storage of prompts, models, and outputs for reproducibility and certification over long lifecycles. &
C2: “Challenge: how to check in the version of the model for future reference.” \newline
C3: “We use GitLab for instruction files; concise generation ensures traceability.” & C2, C3 \\ \hline

\multirow{3}{*}{Theme 3}
& AI Training and Upskilling &
Build AI acceleration or upskilling teams to train engineers on AI tools, compliance, and productivity enhancement. &
C2: “We have acceleration teams: tech scouting, product compliance, and Pentest tools.” \newline
C4: “Need to raise AI awareness through hackathons to spread knowledge.” & C2, C4 \\ \cline{2-5}

& Tool Standardization through MCP Adoption &
Use MCP to standardize tool interactions and ensure secure, auditable communication between AI systems and enterprise data. &
C2: “We're pushing all of our tools have to be in our MCP, have an MCP API as well, so these agents don't have any limitations.” \newline
C4: “Would like to see things coming top-down to get the mandate.” & C2, C4 \\ \cline{2-5}

& Collaborative Experimentation &
Encourage internal experimentation and hackathons to explore AI integration opportunities and cross-team learning. &
C3: “This is a hackathon where a lot of developers used AI, not because they were told to but because it was efficient.” \newline
C1: “Dedicated AI team experimenting and spreading knowledge.” & C1, C3 \\ \hline

\end{tabular}
\label{tab:practices}
\end{table*}

\subsubsection{Theme: Orchestrated AI Workflow}
The first theme captures how embedded software teams are integrating AI capabilities directly into their development lifecycles. These practices reflect a shift from traditional automation toward more intelligent, feedback-driven engineering processes.

The first practice is the \textbf{\textit{Architecture of Agentic Pipeline}}, where organizations are experimenting with \textit{multi-agentic CI/CD pipelines}. In these setups, generative AI agents collaborate across development stages: implementation, compiling, testing, and deployment. One participant explained, \emph{“We are moving towards a multi-agentic way of development. Agents should work in pipelines, like the teams today”} (C2). This approach extends existing CI/CD pipelines by incorporating autonomous components that perform iterative build, test, and validation tasks.

A second practice, \textbf{\textit{AI-Friendly Artifacts}}, focuses on designing code and documentation structures optimized for generative AI interpretation. Teams are creating structured, machine-readable files such as \textit{agent.md} or annotated instruction sets that describe functional intent, timing constraints, and hardware dependencies. As one participant summarized, \emph{“AI-friendly code basically means more structured comments and clear logic”} (C4). These artifacts serve as a communication layer between human developers and AI agents, improving determinism and reducing hallucination in AI-generated code.

The third practice, \textbf{\textit{Compiler-in-the-Loop Feedback}}, introduces a feedback mechanism where AI-generated code is compiled, simulated, or tested automatically, and the results are fed back to the AI system for refinement. This creates a closed-loop environment where AI learning is guided by executable outcomes rather than textual similarity. As one engineer observed, \emph{“If you just give the LLM access to the compiler, it becomes much better”} (C2). Such compiler-grounded feedback aligns AI generation more closely with the strict performance and correctness requirements of embedded systems.

Finally, using \textbf{\textit{AI for Testing and Validation}} captures how teams apply generative AI to automate parts of their verification workflows, including test case generation, compliance checks, and documentation synthesis. Participants described using agents to create regression tests or produce compliance artifacts aligned with domain-specific standards. For instance, \emph{“We use custom agents for unit testing -- it is a strong addition”} (C3). These applications complement, rather than replace, human validation, and they improve efficiency in the early verification stages.

\subsubsection{Theme: Responsible AI Governance}
The second theme highlights how organizations are formalizing governance frameworks to ensure the responsible and transparent integration of generative AI into development workflows. These practices aim to strike a balance between automation and human oversight while maintaining accountability in safety-critical contexts.

The first, and probably the most important, practice is \textbf{\textit{Prompt Management}}, which focuses on making AI prompting systematic, auditable, and reusable. Teams are creating version-controlled prompt repositories that define standardized templates, contextual instructions, and domain-specific parameters. One participant described this evolution: \emph{“We use shared instructions among all developers to ensure consistent styles”} (C3). These repositories establish consistency across projects and allow organizations to review and refine prompts as formal artifacts.

The second practice, \textbf{\textit{Human-in-the-Loop Supervision}}, ensures that engineers remain active decision-makers throughout AI-augmented development. Participants consistently emphasized that AI systems are valuable collaborators but cannot yet operate autonomously in safety-critical environments. As one engineer explained, \emph{“Human-in-the-loop works best when the engineer keeps control of the method”} (C2). This supervisory model maintains human accountability while leveraging the speed and adaptability of AI.

Another essential practice is \textbf{\textit{Governance and Compliance}}, where companies are expanding their Responsible AI policies to include licensing audits, data access controls, and validation processes for AI-generated artifacts. Several organizations have adopted internal approval workflows before integrating AI-generated code into production, often involving compliance teams or “AI champions” responsible for reviewing the ethical and legal implications. As one participant put it, \emph{“We need to know why we are adopting AI”} (C1), reflecting the deliberate and policy-driven nature of the adoption of AI.

The fourth practice, \textbf{\textit{Traceability and Version Control}}, addresses the challenge of reproducing AI outputs over long product lifecycles. Teams are developing traceability frameworks to capture model versions, prompts, and configurations alongside source code repositories. One engineer noted, \emph{“We use GitLab for instruction files; concise generation ensures traceability”} (C3). This integration of generative AI context into software repositories is rapidly turning into a foundational element of governance, particularly in industries with ten-year support or certification requirements. 

\subsubsection{Theme: Sustainable AI Adoption}
The third theme highlights how organizations develop the capacity and infrastructure necessary to support the adoption of generative AI at scale.

The first emerging practice, \textbf{\textit{AI Training and Upskilling}}, reflects a growing recognition that engineers need to complement their technical literacy in generative AI tools with an understanding of responsible use and compliance requirements. Several organizations have established dedicated \textit{AI acceleration teams} to coordinate these efforts. These teams organize internal workshops, mentoring sessions, and cross-functional hackathons to raise awareness and build confidence in using generative AI-augmented tools. As one engineer explained, \emph{“We have acceleration teams for tech scouting, product compliance, and Pentest tools”} (C2). Another participant emphasized the cultural aspect of these initiatives: \emph{“We need to raise generative AI awareness through hackathons to spread knowledge”} (C4). These structured learning opportunities are helping engineers develop the hybrid skills, both domain-specific and AI-related, required for future embedded development.

The second practice, \textbf{\textit{Tool Standardization through Model Context Protocol (MCP) Adoption}}, demonstrates how organizations are formalizing generative AI use through shared interfaces and communication protocols. Participants described ongoing efforts to implement the MCP as a standardized layer for connecting generative AI systems with enterprise data and engineering tools. MCP enables secure, auditable, and interoperable interactions between generative AI agents and internal repositories, ensuring that each tool operates under consistent access and compliance rules. One participant noted that \emph{“We're pushing all of our tools have to be in our MCP, have an MCP API as well, so these agents don't have any limitations”} (C2), highlighting the need for architectural uniformity in multi-agent environments. In this sense, MCP adoption is not only a technical choice but also a vehicle for tool standardization, seen as essential for scaling generative AI solutions from isolated experiments to organization-wide workflows.

The third practice, \textbf{\textit{Collaborative Experimentation}}, reflects how organizations are creating spaces for exploratory, team-driven innovation. Hackathons, pilot projects, and informal cross-team experiments allow developers to test new generative AI tools and share lessons learned across departments. One participant shared, “This is a hackathon where a lot of developers used generative AI, not because they were told to but because it was efficient” (C3). Another added, \emph{“We have a dedicated AI team experimenting and spreading knowledge”} (C1). These initiatives not only accelerate technical discovery but also foster a sense of shared ownership and learning, helping organizations identify viable generative AI applications for embedded contexts.

\subsection{Challenges}
\begin{figure}[htbp]
    \centering
    \includegraphics[width=.95\linewidth]{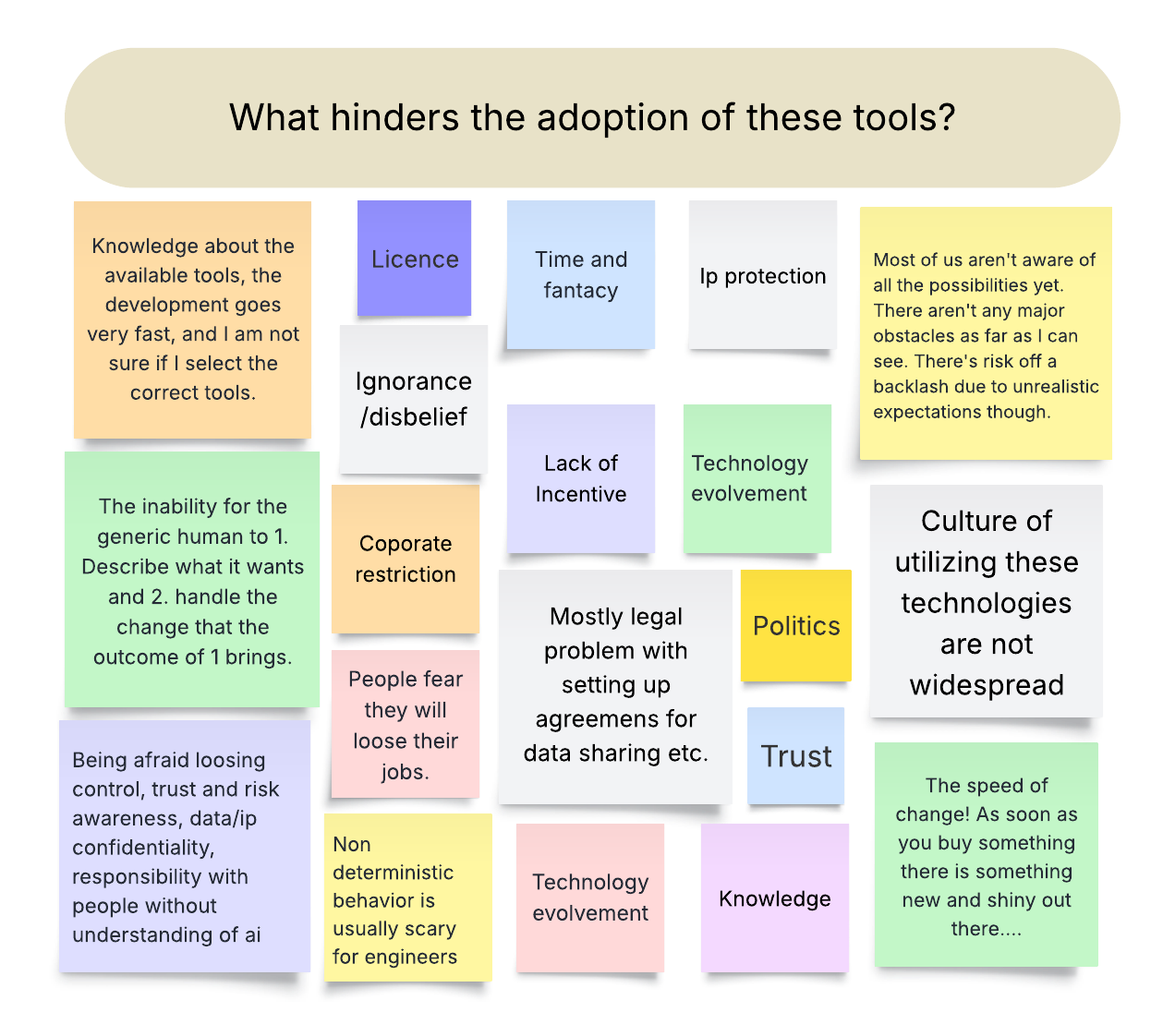}
    \caption{An example question from the Mentimeter-based brainstorming session.}
    \label{fig:menti}
\end{figure}

\noindent
Alongside the emerging practices identified in the study, participants also discussed significant challenges that complicate the adoption of agentic and AI-augmented workflows in embedded software engineering. An example of brainstorming questions and answers is presented in Fig.~\ref{fig:menti}. 

These challenges span the same three thematic areas introduced in the previous section. Table~\ref{tab:challenges} summarizes the key challenges identified within each theme.

\begin{table*}[htbp]
\centering
\caption{Challenges in Adopting Agentic Pipelines for Embedded Systems}
\label{tab:challenges}
\begin{tabular}{p{1cm} p{3.5cm} p{5.5cm} p{5.5cm}}
\hline
\textbf{Theme} &
\textbf{Challenge} &
\textbf{Description} &
\textbf{Evidence from Transcripts} \\ \hline

\multirow{4}{*}{Theme 1} &
Certification gap for AI-generated artifacts &
Lack of established processes and criteria to certify code or artifacts produced by AI agents. &
C2: need to certify agentic systems; questions of code ownership and liability. \\ \cline{2-4}
&
Limited incentives for experimentation &
Performance metrics and delivery pressure discourage developers from exploring AI tools. &
C2: developers hesitant to try AI tools due to productivity metrics. \\ \cline{2-4}
&
Developer resistance and fear &
Skepticism and anxiety about AI’s reliability and its impact on existing roles. &
C1: conservative culture; uncertainty about AI’s impact on roles. \\ \cline{2-4}
&
Re-skilling needs for AI orchestrators &
Existing staff lack the hybrid skills needed to combine domain expertise with AI tooling. &
C1: growing demand for domain expertise over pure programming. \\ \hline

\multirow{5}{*}{Theme 2} &
Unclear ownership and licensing for AI-generated code &
Uncertainty about ownership and license compliance for AI-generated content in products. &
C1: concerns about license violations in LLM-generated code; compliance checks in repositories. \\ \cline{2-4}
&
Non-determinism undermining reproducibility &
Stochastic model behavior makes it hard to guarantee repeatable outputs for certification. &
C2: concerns about verifying outputs and deterministic behavior. \\ \cline{2-4}
&
No robust versioning of prompts, models, and toolchains &
Missing mechanisms to record the exact generative context for each artifact. &
C3: questions about storing prompts or intermediate artifacts for future reuse. \\ \cline{2-4}
&
Traceability challenges over long product lifecycles &
Difficulty maintaining verifiable links between models, prompts, and binaries over 10+ years. &
C2: inability to “check in” LLMs for long-term traceability; automotive safety requirements. \\ \cline{2-4}
&
Dependence on proprietary AI vendors &
Reliance on external platforms for model versions, stability, and access terms. &
C1: concerns over reproducibility and validation of AI-generated tests. \\ \hline

\multirow{5}{*}{Theme 3} &
Uncertain and evolving regulatory requirements &
Rapidly changing AI regulations create ambiguity around compliance obligations. &
C2: EU regulatory boundaries discussed by participants. \\ \cline{2-4}
&
High demands for auditability and explainability &
Internal and external stakeholders require transparent, inspectable AI behavior. &
C4: strict internal policies and blocked access to external AI tools. \\ \cline{2-4}
&
Uneven adoption across teams &
Some teams advance quickly while others lag, creating inconsistent AI maturity. &
C2: acceleration teams and AI training programs; uneven scaling of adoption. \\ \cline{2-4}
&
Siloed organizational structures &
Limited cross-team communication hinders sharing of AI practices and lessons learned. &
C4: large-scale silos and lack of communication between departments. \\ \cline{2-4}
&
Misalignment between leadership and local practices &
Strategic AI ambitions are not always matched by concrete mandates or resources at team level. &
C4: lack of top-down mandates, conflicting goals between units. \\ \hline

\end{tabular}
\end{table*}

\subsubsection{Theme: Orchestrated AI Workflow}
The first group of challenges highlights the technical and procedural complexities of embedding AI systems into existing engineering pipelines.

A primary concern was the \textbf{\textit{certification gap for AI-generated artifacts}}. Participants repeatedly emphasized that while AI-augmented tools can improve productivity, they introduce uncertainty regarding certification and safety assurance. As one engineer noted, “We need to certify agentic systems, but it’s unclear how to do that when models keep changing” (C2). Traditional certification frameworks are built for deterministic systems, leaving little guidance for verifying AI components whose outputs may vary across runs or model versions.

A second challenge relates to \textbf{\textit{limited incentives for experimentation}}. Although participants described several successful pilot projects, organizational performance metrics, such as sprint velocity or feature output, often discouraged exploratory use of AI tools. One participant remarked that “developers are hesitant to try AI tools because it makes them appear less productive” (C2). Without explicit management support, experimentation is often perceived as risky rather than innovative.

A third challenge concerns \textbf{\textit{resistance or fear among developers}}. Some engineers expressed skepticism about AI’s reliability or worried that increased automation might diminish their professional value. A participant from a traditionally conservative team explained, “People are cautious. They are not sure what AI will mean for their role” (C1). Such uncertainty slows down adoption, especially in safety-critical environments where human accountability is essential.

Finally, the ongoing \textbf{\textit{need for re-skilling}} creates friction in daily workflows. Engineers with strong embedded expertise often lack experience with generative AI or data-driven tools, making integration uneven across teams. As one participant summarized, “We need AI orchestrators, but domain experts don’t yet have that training” (C1). These skill gaps not only limit immediate tool usage but also hinder the scalability of AI-augmented workflows.

\subsubsection{Theme: Responsible AI Governance}
The second theme encompasses challenges related to trust, reproducibility, and the broader accountability structures surrounding AI-generated outputs.

A major challenge is the \textbf{\textit{unclear ownership and licensing for AI-generated code}}. Participants expressed uncertainty about intellectual property rights and the legality of using AI-generated content in commercial products. One engineer explained, “We have to be careful because we don’t know who owns the generated code” (C1). This lack of legal clarity creates hesitation to deploy AI-generated code in production environments.

Another recurring issue is the \textbf{\textit{non-deterministic behavior of LLMs}}. Participants observed that identical prompts may yield different results across runs, making verification difficult. As one participant explained, “We cannot guarantee the same output each time, which complicates reproducibility” (C2). For embedded systems, where timing, resource constraints, and deterministic execution are essential, this unpredictability poses a significant validation challenge.

Relatedly, several engineers highlighted the \textbf{\textit{absence of robust versioning mechanisms}} for prompts, model configurations, and supporting data. Without these, it becomes nearly impossible to reproduce the exact generative context for a given software artifact. “We need to store prompts and model hashes, but there’s no process for that yet” (C3). This gap undermines long-term maintainability, especially in domains requiring decades of traceability.

Participants also described difficulties in \textbf{\textit{maintaining traceability over long product lifecycles}}. For industries like automotive, where systems may need to be maintained for ten years or more, the continuous evolution of models introduces version drift. One participant explained, “We can check in the compiler from ten years ago, but not the LLM” (C2). The inability to archive AI models in a certifiable manner complicates future audits and regulatory compliance.

Lastly, participants warned about \textbf{\textit{dependence on proprietary vendors}}. Many organizations rely on third-party AI platforms, making them vulnerable to external updates, license changes, or deprecations. As one engineer expressed, “We depend on vendors for stability, and they can change the model or access terms anytime” (C1). This dependency raises questions about long-term sustainability and control over critical development infrastructure.

\subsubsection{Theme: Sustainable AI Adoption}
The third theme encompasses challenges at the organizational and regulatory levels that influence how AI tools are adopted, scaled, and governed.
One pressing issue is the \textbf{\textit{impact of evolving regulatory obligations}}, particularly in relation to the EU~AI~Act and other emerging standards. Participants noted that the pace of AI innovation often exceeds the development of legal frameworks, leaving companies uncertain about compliance boundaries. One participant explained, “We are waiting for clear guidance on what the AI Act will mean for software validation” (C2). This regulatory uncertainty creates hesitation in fully deploying AI systems within embedded contexts.

Another challenge involves \textbf{\textit{auditability and explainability}}. Participants described internal policies that restrict or delay access to AI tools due to security and traceability concerns. One engineer mentioned, “We have strict internal policies and blocked access to external AI tools” (C4). Organizations must balance the desire for innovation with the need to ensure transparency and accountability.

At the cultural level, \textbf{\textit{uneven adoption across teams}} remains a consistent challenge. Some groups have fully embraced AI-augmented development, while others remain skeptical or resource-constrained. As one engineer summarized, “Some parts of the organization are far ahead, others are still blocked by IT policies” (C2). These disparities result in inconsistent maturity across projects and hinder the creation of a unified AI adoption strategy.

\textbf{\textit{Siloed organizational structures}} further amplify these inconsistencies. Participants reported limited cross-departmental communication about AI initiatives, making it difficult to share best practices or develop standardized workflows. “Different departments are experimenting on their own,” one participant explained, “and we don’t always know what others are learning” (C4).

Finally, \textbf{\textit{misalignment between leadership vision and local practices}} was frequently cited as a barrier. While executives may prioritize AI adoption as a strategic goal, middle management and engineering teams often lack clear guidance or resources. A participant noted, “Leadership wants AI adoption, but we don’t have the mandate or support to implement it effectively” (C4). This disconnect slows organizational momentum and undermines coordinated adoption efforts.

Across all three themes, participants identified that while enthusiasm for AI integration is high, significant challenges remain in realizing sustainable, responsible, and standardized adoption.
Technical barriers such as certification and determinism (Theme 1), governance issues around accountability and reproducibility (Theme 2), and organizational constraints related to regulation, structure, and leadership alignment (Theme 3) collectively illustrate the multi-dimensional nature of adopting AI in embedded software engineering.

\section{Related Work}
\noindent
Studies on the intersection of artificial intelligence and software engineering has evolved rapidly over the past decade, moving from traditional ML integration to foundation-model-based software systems and LLMs. Early work on machine learning for software engineering explored the distinct challenges of integrating ML components into production systems, such as data versioning, model drift, and testing pipelines~\cite{nahar2023meta, amershi2019software, zhang2020machine}. These studies established that developing ML-enabled software requires specialized practices distinct from conventional engineering workflows. Subsequent research has investigated the unique disruptions caused by large, pre-trained language models. These efforts~\cite{hassan2024rethinking, parnin2025building, nahar2025beyond, terragni2025future} document issues including non-deterministic outputs, difficulties in defining correctness, and the lack of systematic evaluation or governance structures.

Hassan et al.~\cite{hassan2024rethinking} propose a theoretical framework for what they call FMware: software systems built on foundation models. Drawing on industrial experience and research synthesis, they identify ten major challenges in developing trustworthy FMware, including prompt fragility, data alignment, and multi-generational software integration. Their framework formalizes a new lifecycle for foundation-model software, encompassing cognitive architecture design, workflow orchestration, and trustworthiness engineering, and calls for new principles of reproducibility and semantic observability across AI generations.

Parnin et al.~\cite{parnin2025building} conduct one of the first empirical investigations of professional software engineers building AI-enabled “copilot” products. Through interviews and structured brainstorming sessions, they identify pain points throughout the lifecycle: prompt engineering, orchestration, testing, and compliance, and highlight the absence of best practices and tool support. Their findings reveal that most practitioners approach prompt development through trial and error, underscoring the need to “add engineering to prompting” by introducing validation and debugging processes for prompts. This work helps frame prompt authoring and lifecycle management as new software-engineering competencies.

Nahar et al.~\cite{nahar2025beyond} extend this line of inquiry with a large-scale mixed-methods study involving 26 interviews and a survey of 182 Microsoft practitioners developing LLM-based products. Their study confirms previously reported challenges: such as ambiguous specifications, testing subjectivity, and unreliable validation pipelines, and catalogs 19 emerging practices to address them. These include combining qualitative and quantitative metrics for evaluation, using LLMs as judges, and automating offline testing. Their findings provide one of the most comprehensive empirical accounts of how industrial teams are adapting workflows and quality-assurance processes to LLM-driven disruption.

Terragni et al.~\cite{terragni2025future} present a complementary vision of the future of AI-driven software engineering, describing a symbiotic partnership between human engineers and specialized AI agents across the entire software lifecycle. Their work conceptualizes software engineering as evolving from traditional code-centric development to agentic and collaborative AI systems. They foresee the emergence of multi-agent environments that augment human creativity, validation, and testing, laying out a research agenda for trustworthy and transparent human–AI co-development.

Beyond these major contributions, earlier design-oriented studies explore how developers interact with LLM-powered tools. Vaithilingam et al.~\cite{vaithilingam2022expectation} examine user experiences with AI-augmented code completion in Visual Studio, while Murali et al.~\cite{murali2023codecompose} analyze the deployment of Meta’s CodeCompose system, revealing concerns around trust, evaluation metrics, and workflow adaptation. Jiang et al.~\cite{jiang2022promptmaker} and Brade et al.~\cite{brade2023promptify} introduce prototyping tools such as PromptMaker and Promptify to support prompt authoring and iterative refinement, and Liang et al.~\cite{liang2025prompts} emphasize that prompts are part of programs as well, highlighting the growing need for structured interfaces and evaluation pipelines for prompt-based systems. The Model Context Protocol (MCP)~\cite{Anthropic2024ModelContext} provides an open, vendor-neutral standard for linking LLMs and AI agents to organizational systems securely and transparently.

Collectively, these works illustrate a rapid transition from studying traditional ML integration to exploring generative and agentic software engineering paradigms. While prior research provides both conceptual visions~\cite{hassan2024rethinking,terragni2025future} and empirical insights from product teams~\cite{parnin2025building, nahar2025beyond}, most efforts focus on general-purpose or enterprise software. In contrast, the domain of embedded systems, where determinism, safety, and long-term traceability are paramount, remains underexplored. This paper addresses that gap by examining how organizations in the embedded domain are integrating generative AI responsibly, focusing on strategies, protocols, and human–AI collaboration mechanisms for sustainable and auditable adoption.

\section{Implications}
~\label{sec:discussion}
The findings from our study reveal that integrating generative AI into embedded software development is not merely a technical challenge but a multi-dimensional transformation that touches strategy, infrastructure, and professional identity. Rather than simply introducing new tools, agentic pipelines are reshaping how organizations distribute responsibility, interpret reproducibility, and define trust in automated systems.

We categorize possible implications into three themes: (1) emerging protocols, (2) strategic approaches, and (3) future and broader impacts.

\subsection{Emerging protocols}

\subsubsection{AI-Friendly Code Protocol (AICP)}
While the \textit{MCP} provides the infrastructure to exchange these artifacts securely between AI systems and enterprise data sources, the \textit{AICP} defines how information within a software project should be structured for AI interpretation. MCP enables two-way, authenticated communication that enforces access control and auditability, addressing organizational concerns about data governance and model transparency.  
As one participant explained,\emph{“All tools should have an MCP-compliant API so the agents can use them safely inside the pipeline.”}
Together, AICP and MCP form complementary layers in an agentic CI/CD pipeline: AICP governs \textit{what} is exchanged and how it is represented, while MCP governs \textit{how} it is transmitted, validated, and logged. This combination supports a secure, traceable, and interoperable foundation for enterprise-grade integration of AI into embedded workflows.

\subsubsection{AI-Friendly Code}
Participants widely agreed that structured, context-rich, and machine-readable code -- termed \textit{AI-friendly code}, is essential for effective collaboration with generative agents and therefore a key to realize AICP. Some viewed this as a temporary adaptation while models mature; others argued that it represents a durable evolution of good engineering hygiene.  
In embedded systems, where determinism and traceability are critical, such practices are likely to persist. Embedding metadata, model identifiers, and generation context within code artifacts supports both machine comprehension and long-term maintenance, extending documentation standards into the era of human–AI co-development.

\subsubsection{Prompts as First-Class Artifacts}
Participants consistently emphasized that prompts and instruction files must be treated as formal project assets rather than ephemeral text inputs. They capture design intent, constraints, and reasoning, providing essential context for reproducing or auditing generated code.\emph{“We started saving every prompt we used for production code -- it’s part of our documentation now.”}
Versioning prompts alongside source code not only enables reproducibility but also creates a new traceable link between human intent and AI-generated artifacts, forming the foundation of AI-aware configuration management.

\subsection{Strategic Approaches}

\subsubsection{Top-Down vs. Bottom-Up Adoption}
Organizations differ widely in how they approach AI adoption. Two primary trajectories emerged: \textbf{top-down} strategies, driven by leadership mandates and centralized infrastructure, and \textbf{bottom-up} strategies, led by practitioner experimentation. Each path offers distinct advantages and risks. Top-down approaches ensure alignment, compliance, and resource allocation but can suppress local innovation. Bottom-up adoption, in contrast, fosters creativity and rapid learning but may result in fragmented practices and governance gaps.  
One engineer summarized the tension succinctly:\emph{“We have the push from management to use AI, but in practice, it’s the small teams experimenting that really learn how it works.”}
A hybrid strategy, where leadership defines boundaries and incentives, while teams locally adapt and iterate, appears to offer the most sustainable path for scaling agentic workflows in complex embedded environments.

\subsubsection{What Counts as Reproducible with LLMs?}
Reproducibility is a defining concern for embedded development, where identical toolchains are expected to yield identical binaries. Generative systems disrupt this expectation because LLM outputs are inherently stochastic and evolve over time. Participants emphasized the need for \textit{traceability of the generative context}, including prompts, model versions, and configurations.  
As one participant noted,\emph{“In automotive, we can rebuild the whole environment from fifteen years ago, but we can’t check in a model the same way.”}
Another added,\emph{“We need to store not just the code but how it was generated -- the prompt, the model, everything.”}
These concerns suggest that reproducibility may need to be redefined as \textit{functional equivalence under fixed constraints} rather than byte-level identity, aligning with safety certification practices in other regulated domains.

\subsection{Future Directions and Expanding Boundaries}

\subsubsection{When Is the Human-in-the-Loop Skippable?}
Determining when human validation can be safely reduced remains a key open question. Participants agreed that human review is indispensable for safety- and compliance-critical software, yet partial automation is feasible for low-risk tasks such as documentation, test generation, or non-production builds.  
One engineer explained,\emph{“We trust compilers to optimize code without checking every instruction -- maybe one day we’ll trust AI agents the same way.”}
This suggests a \textit{risk-tiered autonomy model}, in which oversight intensity varies by artifact type and by the proven reliability of the AI agent. Over time, as model verification techniques mature, agentic systems could achieve certified autonomy similar to compilers or testing frameworks today.

\subsubsection{Beyond Software Engineering}
Although this study focused on software development, participants observed and anticipated that generative AI and agentic systems would extend far beyond engineering roles. Several mentioned parallel transformations in corporate functions such as finance, human resources, and operations, where AI agents are already automating documentation, analytics, and decision workflows -- creating measurable business value worth millions of dollars.\emph{
“It’s not just developers -- AI will change how everyone works, from engineers to accountants.”}
These observations suggest that the integration of AI into embedded engineering may serve as a model for broader organizational adaptation, requiring coordinated strategies for training, governance, and ethical oversight across departments.

\subsubsection{Is It Too Late to Start Exploring AI?}

Not all participating organizations have yet integrated AI into their production workflows. Several participants expressed uncertainty about whether their current level of preparedness, expertise, or strategic alignment would affect their competitiveness in the coming years. One participant asked pointedly,\emph{“Are we already too late? The big players have been training their own models for years while we’re still figuring out how to use Copilot.”}

Despite these concerns, the consensus across discussions was that it is never too late to begin exploring AI adoption. While smaller or more traditional companies may have lost the early advantage of developing proprietary foundation models, they can still leverage their unique assets, such as domain-specific data, systems knowledge, and existing processes, to fine-tune or customize existing models. They may not build the models ourselves, but we own the data. Fine-tuning on what they already know could be their shortcut.”

Participants also observed that late adopters might even benefit from entering the field at a more mature stage, bypassing the experimental pitfalls of early implementations and capitalizing on more stable toolchains, APIs, and governance frameworks. For embedded system providers, this approach could translate into faster, safer integration of generative AI, using the latest agentic architectures and context protocols without the technical debt of legacy experiments.

In summary, participants envisioned a near future where agentic CI/CD pipelines reshape embedded software engineering through structured protocols, standardized artifacts, and evolving human–AI collaboration models. Achieving this vision will require balancing strategic alignment, robust governance, and cultural readiness to ensure that automation enhances not compromises the rigor and reliability that embedded systems demand.

\section{Conclusion}
~\label{sec:conclusion}
\noindent
This study investigated how organizations in the embedded software domain are adopting and adapting to generative AI, and what this means for engineering practices, professional roles, and governance. Through qualitative analysis of focus group interviews and brainstorming sessions with ten senior engineers across four companies, we identified eleven emerging practices and fourteen key challenges, clustered into three overarching themes: Orchestrated AI Workflow, Responsible AI Governance, and Organizational Standardization.

Embedded software teams are beginning to integrate AI into CI/CD pipelines, design AI-friendly artifacts, formalize prompt management and human-in-the-loop supervision, and build institutional capacity through training, standardization, and the use of protocols such as MCP. At the same time, they face challenges related to certification of AI-generated artifacts, reproducibility in the presence of non-deterministic models, long-term traceability, and organizational alignment across silos and regulatory constraints.

Taken together, these findings suggest that adopting generative AI in embedded development is not only a matter of tool integration but also one of reshaping workflows, governance models, and competency profiles. Successful adoption will require embedded organizations to co-evolve technical infrastructures and organizational practices so that automation enhances, rather than undermines, the reliability, traceability, and regulatory compliance that their systems demand.

\section*{Acknowledgment}

This work is funded by Software Center, a collaboration between University of Gothenburg, Chalmers and 18 universities and companies -- \url{www.software-center.se}.

\bibliographystyle{IEEEtran}
\bibliography{references}

\end{document}